# A smartphone-based simple method for determination of the free space permeability

Sanjoy Kumar Pal[1,4], Soumen Sarkar[2,4], and Pradipta Panchadhyayee[3,4*]

[1]Anandapur H.S. School, Anandapur, Paschim Medinipur, West Bengal, India
[2]Karui P.C. High School, Hooghly, West Bengal, India
[3]Department of Physics (UG & PG), Prabhat Kumar College, Contai, Purba Medinipur, India
[4]Institute of Astronomy, Space and Earth Science, Kolkata -700054, W. B., India
[*]E-mail: ppcontai@gmail.com

## Abstract

A simple and novel method is designed to determine the free space permeability. This value is computed from the expression of the terminal velocity of a magnet falling through a conducting pipe using the magnetic sensor of a smartphone and a video player. This method deserves its importance because of the accuracy and precision of the results.

## Introduction

Determining free space permeability is essential in various fields, such as civil engineering, environmental science, and agriculture. Traditional methods often require expensive equipment and complex setups, limiting accessibility for many users. However, with the widespread availability of smartphones equipped with various sensors and computing capabilities, conducting simple experiments to measure permeability has become feasible. There has been a growing trend in incorporating smartphone sensors into introductory and first-year university physics courses in recent years. This trend offers exciting prospects for hands-on experimentation and applying physics principles to real-world scenarios [1-4]. As we have aimed at determining the free space permeability via the measurement of the terminal velocity of a magnet when falling through a non-ferromagnetic conducting pipe, we have found several smartphone-sensor-based interesting methods [5-8] reported to measure the terminal velocity of a magnet in the said condition. We have extensively designed a novel way to explore smartphone magnetic sensors in our present study. The experiment also involves tracking the motion of the freely falling magnetic ball inside a conducting pipe. The precision in measurement is enhanced by the processes of tracking and analyzing videos of the measurement setup.

## Theoretical framework

In this work, we have determined the terminal velocity of a magnetic ball of mass ($M$) as it falls under gravity through a conducting pipe. For the motion of the magnet, it creates a changing magnetic field in the pipe, which induces eddy currents. These currents counteract the motion of the magnet. This electromagnetic interaction between the eddy currents and the falling magnet creates an electromagnetic damping force that opposes its free-falling motion under gravity and slows the magnet's fall. Consequently, as the magnet falls, it encounters a viscous medium within the pipe. In this environment, the instantaneous damping force ($F_{\mathrm{em}}$) is directly proportional to the instantaneous velocity ($v$) of the magnet.

$$F_{\mathrm{em}} = -kv. \tag{1}$$

Here, the expression of the proportionality constant $k$ [7] is

$$k = \left(\frac{15}{1024}\right)\mu_0^2 m^2 \sigma \left(\frac{1}{a^3} - \frac{1}{b^3}\right), \tag{2}$$

Here, $\sigma$ represents the conductivity of the material composing the pipe. Parameters $a$ and $b$ denote the inner and outer radii of the pipe, respectively. Additionally, $\mu_0$ represents the permeability of free space, while $m$ stands for the magnetic moment of the magnet.

Now, we can write

$$kv_T = Mg. \tag{3}$$

After a considerable time, though it is very small, the damping force becomes equal to the weight of the falling magnet. Eventually, the velocity of the magnet reaches its terminal velocity ($v_T$). We neglect the viscous force and buoyant force due to air. Measuring $v_T$ we can easily estimate the value of $k$ using the values of the mass of the magnetic ball and the acceleration due to gravity in Eq. (3).

**Experimental Results**

Here, we have used a ball magnet with a diameter of 7.69 mm and mass of 1.821 g and a copper pipe with an inner and outer diameter of 9.56 mm and 15.86 mm, respectively, and a length of 101.0 cm. For measuring diameters and mass, we use digital slide callipers (vernier constant 0.01 mm) and a digital scale balance (0.001 g accuracy).

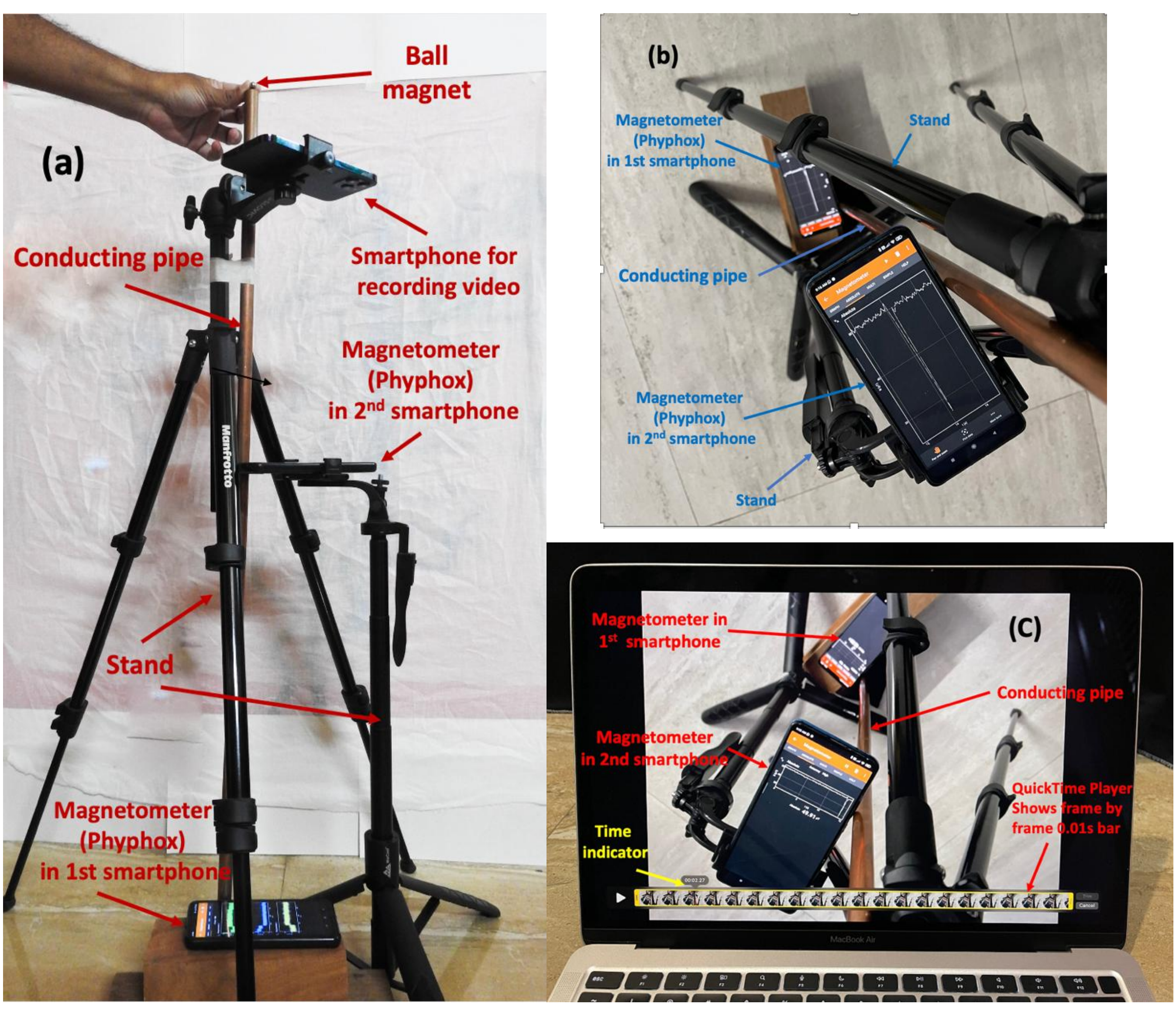

Fig. 1: (a) Experimental setup, (b) Setting of two smartphones, (c) Analyzing the recorded video.

In the first case, the pipe is kept vertically, allowing a particular ball magnet to fall through it. We positioned two smartphones near the pipe at distances of 35.0 cm and 87.0 cm from the top of the pipe. These smartphones were set to the phyphox magnetometer mode 'on' at their respective locations (Refer to Fig. 1a). We have measured the widths (9.6 mm for the first one and 9.8 mm for the second one) of both smartphones separately using a slide callipers. We have marked dots at the midpoints of the width of each phone with a marker pen and measured the distance between the dots of the phone using a standard wooden metre scale.

After setting the above arrangement, the ball magnet is allowed to fall through the pipe. We have examined the ball magnet, and it is found that it achieves terminal velocity shortly after being released into the pipe, having traversed a minimal distance. As the ball magnet falls through a pipe, the magnetometer in the phyphox app captures the magnetic field's temporal variation. The magnetic field values exhibit fluctuations, increasing when the magnet is in proximity and decreasing as it moves away. The magnetometer waveforms reveal distinct, sharp peaks in the waveforms of the magnetic field variation profile. To calculate the terminal velocity of the ball magnet, we require the time interval corresponding to the occurrence of peaks in the two smartphones positioned at different locations along the pipe (Refer to Fig. 1b). To accomplish this, a short video capturing the entire system, along with the magnetometer waveforms from the two smartphones, was recorded using another smartphone (iPhone 12 Pro Max) with a frame rate of 60 frames per second. Subsequently, the recorded video file was transferred to a MacBook Pro, where the time interval was accurately measured (with a minimum precision of 0.01s) using the QuickTime Player, the built-in player of the MacBook (See Fig. 1c). To display the millisecond time bar while playing a video on QuickTime Player one must press the keys (Command + T). For the Windows OS, one can use a VLC media player with the 'Time v3.2' extension. Using the method discussed above, we successfully calculated the terminal velocity. We replicated the experiment, collecting data for four additional sets with the same ball magnet, wherein the upper smartphone was incrementally lowered by approximately 4 to 6 cm each time. Table 1 summarizes all the collected data, revealing a high degree of consistency in the calculated values for the terminal velocity of the ball magnet. Once the terminal velocity is determined, we proceed to calculate the value of $k$ (refer to Eq. 3). As we know that the conductivity of copper at $25^0$C is $5.6754\times10^7$ $(\Omega m)^{-1}$, and the gravitational acceleration is 9.8 $ms^{-2}$, we can derive the value of the permeability of free space, $\mu_0$ , knowing the value of magnetic moment ($m$) of the used magnet. Utilizing the standard torsional method, we have measured the magnetic moment of the ball magnet as 0.1774 $Am^2$.

Table 1: Table for the measurement of terminal velocity and permeability of free space

| Distance between two smartphones (m) | Time interval between two peaks (s) | Terminal velocity, $v_T$ (ms$^{-1}$) | $k = \frac{Mg}{v_T}$ (kg.s$^{-1}$) | $\mu_0$ ($\pi\times10^{-7}$ Hm$^{-1}$) | Average value of $\mu_0$ ($\pi\times10^{-7}$ Hm$^{-1}$) |
|---|---|---|---|---|---|
| 0.520 | 8.33 | 0.0624 | 0.2918 | 3.98 | 3.99 |
| | 8.40 | 0.0619 | 0.2943 | 3.99 | |
| | 8.38 | 0.0621 | 0.2936 | 3.99 | |
| 0.472 | 7.56 | 0.0624 | 0.2918 | 3.98 | |
| | 7.59 | 0.0622 | 0.2929 | 3.99 | |
| | 7.63 | 0.0619 | 0.2945 | 4.00 | |
| 0.425 | 6.84 | 0.0621 | 0.2932 | 3.99 | |
| | 6.88 | 0.0618 | 0.2949 | 4.00 | |
| | 6.86 | 0.0620 | 0.2940 | 3.99 | |
| 0.381 | 6.12 | 0.0623 | 0.2926 | 3.98 | |
| | 6.16 | 0.0619 | 0.2945 | 4.00 | |
| | 6.14 | 0.0621 | 0.2936 | 3.99 | |
| 0.322 | 5.20 | 0.0619 | 0.2942 | 3.99 | |
| | 5.17 | 0.0623 | 0.2925 | 3.98 | |
| | 5.22 | 0.0617 | 0.2953 | 4.00 | |

**Conclusion:**

In this work, we have shown the use of technology readily available in smartphones, with a neodymium ball magnet and a non-ferromagnetic conducting pipe for determining the value of free space permeability. The main challenge lies in the high-precision measurement of the terminal velocity. The average value terminal velocity is (0.0621 ± 0.0002 ± 0.0001) ms$^{-1}$. Here, the instrumental uncertainty and statistical error are given successively with the mean value of terminal velocity. In the same fashion, we present the mean value of the magnetic moment of the ball magnet as (0.1774 ± 0.0005 ± 0.0002) Am$^2$. Considering the calculations on errors, the obtained value of free space permeability (in $\pi\times10^{-7}$ Hm$^{-1}$ unit) is 3.99 ± 0.01 ± 0.002. Finally, the work highlights that the combined use of smartphone sensors and simple video analysis enhances the precision of the measurements.